\documentclass[11pt,article]{article}
\usepackage{graphicx}
\usepackage{indentfirst,csquotes}

\usepackage{fancyhdr,etoolbox}

\usepackage{color}
\usepackage{soul}
\usepackage{multirow}
\usepackage{algorithmic}
\usepackage{upgreek}
\usepackage{tabulary}
\usepackage{textcomp}
\usepackage{amsmath,amsfonts}
\usepackage{algorithm}
\usepackage{makecell}

\usepackage[caption=false, font=footnotesize]{subfig}
\usepackage{mathtools}

\usepackage{amsthm}
\usepackage{xcolor}

\usepackage[inline]{enumitem}

\usepackage{multicol}
\usepackage{dsfont}
\usepackage{lipsum}

\DeclareMathAlphabet{\bi}{OML}{cmm}{b}{it}
\DeclareMathAlphabet{\bcal}{OMS}{cmsy}{b}{n}
\DeclareMathAlphabet{\brmn}{OT1}{cmr}{bx}{n}

\DeclareMathSymbol{\R}{\mathalpha}{AMSb}{"52}

\newcommand{\bfeta}{\boldsymbol{\eta}}

\newcommand{\Tau}{\boldsymbol{\mathcal{T}}}

\def\x{\mathbf{x}}

\def\J{{\cal J}}
\def\N{{\cal N}}

\def\bA{\mathbf{A}}

\def \a{\mathbf{a}}
\def \v{\mathbf{v}}
\def \y{\mathbf{y}}
\def \z{\mathbf{z}}

\def \w{\mathbf{w}}
\def \p{\mathbf{p}}

\def \q{\mathbf{q}}

\def \Real{\mathbb{R}}
\def \complex{\mathbb{C}}

\title{Robust Phaseless Imaging via\\ Reverse Kullback-Leibler Divergence} 

\date{}

\author{Nazia Afroz Choudhury, Bariscan Yonel, Birsen Yazici}

\begin{document}

\maketitle

\begin{abstract}
Robustness to noise and outliers is a desirable trait in phase retrieval algorithms for many applications in imaging and signal processing. In this paper, we develop novel robust phase retrieval algorithms based on the minimization of reverse Kullback-Leibler divergence (RKLD) within the Wirtinger Flow (WF) framework. We use RKLD over intensity-only measurements in two distinct ways: \emph{i}) to design a novel initial estimate based on minimum distortion design of spectral estimates, and \emph{ii}) as a loss function for iterative refinement based on WF. The RKLD-based loss function offers implicit regularization by processing data at the logarithmic scale and provides the following benefits: suppressing the influence of outliers and promoting projections orthogonal to noise subspace. We perform a quantitative analysis demonstrating the robustness of RKLD-based minimization as compared to that of the $\ell_2$ and Poisson loss-based minimization.
We present three algorithms based on RKLD minimization, including two with truncation schemes to enhance the robustness to significant contamination. Our numerical study uses data generated based on synthetic coded diffraction patterns and real optical imaging data. The results demonstrate the advantages of our algorithms in terms of sample efficiency, convergence speed, and robustness with respect to outliers over the state-of-the-art techniques.
\end{abstract} 

\section{Introduction}
\subsection{Motivation and Overview of Our Approach}
Phase retrieval \cite{fienup2013phase,klibanov1995phase} is a problem of great interest in signal processing\cite{balan2006signal, Shechtman2014} and imaging, including diffraction imaging\cite{candes2015phase2}, X-ray crystallography\cite{millane1990phase}, Fourier optical imaging\cite{shechtman2015phase}. The problem refers to the recovery of a real or complex unknown signal from its intensity only measurements. In many applications, the measurements are modeled as
\begin{equation}\label{eq:phaless}
y_m = | \langle \a_m, \x \rangle |^2, \quad m = 1, 2, \cdots M
\end{equation}
where $\x \in \complex^N$ is the unknown signal and $\a_m \in \mathbb{C}^N$ is the $m$-th sampling or measurement vector. Thus, the phase retrieval problem requires solving a system of $M$ quadratic equations and, in general, is known to be NP-hard\cite{candes2015phase}.

In many practical imaging applications, the measurements are contaminated with noise and outliers\cite{weller2015undersampled, li2022poisson}. 
Conventionally, the choice of a particular loss function for an inference problem is motivated through the underlying noise distribution in the measurement process\cite{li2022poisson,deming2007phase}. However, in many inverse problems, the noise distribution may be complex due to multiple sources of corruption, such as measurement error, thermal noise, and photon noise \cite{deming2007phase}. Additionally, the inference may be further complicated by the model errors stemming from the linearization of physics-based modeling, numerical errors in computational models, or grid mismatch effects in the discretization of inverse problems. 

In the context of robust phaseless imaging, we study reverse Kullback-Leibler Divergence (RKLD) minimization, an information-theoretic measure of dissimilarity, from the rationale of having strictly non-negative measurements and the logarithmic processing offered by RKLD. 
Due to the strict non-negativity of the search in the range of the model in \eqref{eq:phaless}, we approach the problem from the perspective of a probability density estimation procedure, where RKLD minimization becomes similar to the variational inference technique.
Most importantly, RKLD minimization generates robust estimates in the presence of outliers due to its logarithmic processing of the measurements, because logarithmic processing \emph{i}) suppresses the influence of large magnitudes in residuals and \emph{ii}) implicitly promotes projections orthogonal to the noise subspace. Taking advantage of these properties, we introduce a class of novel, robust algorithms based on the minimization of RKLD to address the phase retrieval problem in the presence of noise and outliers.

In our approach, we use the Wirtinger Flow (WF) framework\cite{candes2015phase} to minimize the RKLD-loss and
develop three different algorithms for phase retrieval, namely the RKLD-based Wirtinger Flow (RKLD-WF), the median truncated Wirtinger Flow (RKLD-MTWF), and the gradient truncated Wirtinger Flow (RKLD-GTWF).
All these algorithms use the RKLD-based spectral initialization based on the method we developed in \cite{yonel2020spectral}.
Additionally, we investigate the robustness of RKLD with respect to two outlier models using the influence curve analysis\cite{hampel1974influence}. We also study the orthogonality promoting properties of RKLD for noisy measurements. Beyond the characteristic robustness of RKLD, we enhance the RKLD minimization by adapting the sample truncation schemes used in the phase retrieval literature commonly studied with the Gaussian sampling model. Finally, we present numerical simulations using synthetic coded diffraction patterns (CDP)\cite{candes2015phase2}, which is a physically realistic set-up, and real optical data\cite{metzler2017coherent} to compare the performance of the RKLD-loss with those of $\ell_2$ and Poisson-loss as well as other state-of-the-art algorithms.
Our experimental results demonstrate improvements over the state-of-the-art in terms of robustness to additive arbitrary noise, outliers, sample efficiency, and convergence speed. Specifically, the RKLD-based algorithms perform efficiently in challenging conditions such as scarcity of measurements, and high percentage and magnitude of outliers.

\subsection{Related Work and Advantages of Our Approach}

Early studies of phase retrieval developed heuristic error reduction algorithms based on alternating projection methods \cite{gerchberg1972practical, fienup1978reconstruction, fienup1982phase}. In the last decade, several convex and non-convex algorithms, with exact recovery guarantees, 
have been proposed \cite{Candes2013phaselift, candes2015phase2, Waldspurger2015, goldstein2018phasemax, candes2015phase, Candes13a, candes2014solving,Demanet14, hand2016elementary, hand2016corruption, hand2017phaselift}. Most commonly, convex approaches rely on semidefinite relaxation to recover the unknown vector in the form of a rank-one positive definite matrix in the lifted domain. Despite yielding robust estimates, operation in the lifted domain causes these methods to have a high computational and memory cost.
Recently, computationally efficient non-convex approaches, specifically WF\cite{candes2015phase} and its variants \cite{chen2017solving,zhang2016reshaped,zhang2018median,wang2017solving,chen2017robust,wang2018phase,kolte2016phase, Zhang2017b, wang2017staf,li2021poisson}, have gained popularity. The seminal WF algorithm \cite{candes2015phase} solves the following problem based on $\ell_2$-loss function:
\begin{equation}\label{eq:ell_2}
    \min_{\z \in \complex^{N}} \J_{\ell_2}(\z):=\frac{1}{2M}\sum_{m=1}^{M}(|\langle\a_m,\z\rangle|^2 - y_m)^2. 
\end{equation}

Despite the non-convexity of \eqref{eq:ell_2}, the WF-based algorithms successfully deploy a gradient descent-based procedure by fielding an initial estimate sufficiently close to the true solution based on the classical spectral methods \cite{netrapalli2013phase, goldstein2018phasemax, yonel2020deterministic}. However, in the presence of noise and outliers, the local minima deviates from the true solution.
The WF-inspired variants control the impact of the additive outliers\cite{zhang2018median, chen2017robust} as well as the statistical ones\cite{chen2017solving,zhang2016reshaped,wang2017solving,wang2018phase}, arising from the sampling vectors and model mismatch.
To this end, the existing algorithms can be grouped under two major categories: \emph{i}) sample truncation methods to eliminate the contribution of the outliers due to model mismatch and the additive outliers\cite{chen2017solving, zhang2018median, wang2017solving, kolte2016phase}, and \emph{ii}) ``reshaping" of the least-squares formulation by using the amplitude measurements, i.e., $|\langle\a_m,\x\rangle|$ to suppress the outliers in the data arising from the sampling vectors\cite{zhang2016reshaped, zhang2018median, wang2017solving, wang2018phase, chen2017robust}. However, the amplitude-based least-squares formulation is not smooth and susceptible to converging to local minima as observed in \cite{luo2020phase}.


It is well-known that a fundamental limitation of least-squares regression, which the standard WF algorithm is based on, is its susceptibility to biasing estimates towards outliers\cite{huber2004robust}. 
Even with ideal data satisfying \eqref{eq:phaless}, this susceptibility has been observed as a source of degradation in computational and sample efficiency of the WF method under the Gaussian sampling model due to the presence of statistical outliers in the measurement vectors \cite{candes2015phase}. Subsequently, the Poisson-loss function, equivalent to the forward KL divergence (FKLD), is studied within the WF framework \cite{chen2017solving}. However, for $\x \in \Real^N$ and real Gaussian measurement vectors, the gradient of the Poisson loss may become too large due to small values of $|\langle\a_m,\z\rangle|$'s in the denominator and, as a result, the iterates may leave the basin of attraction around the true solution. Therefore, a sample truncation method is deployed to eliminate the unfavorable measurement vectors. However, truncating small $|\langle\a_m,\z\rangle|$'s prevents this method from promoting projections orthogonal to noise subspace; hence, the statistical bias of the gradient increases. 

Unlike Poisson loss-based approach \cite{chen2017solving} and the standard WF algorithm \cite{candes2015phase}, RKLD-based minimization yields stable gradients without the requirement of sample truncation for real-valued data. The gradient of the RKLD-based loss function processes the observed and synthesized intensities on a logarithmic scale, thereby suppressing the effect of large magnitudes. We investigate this characteristic robustness of RKLD by deriving its influence curve\cite{hampel1974influence}, which shows that the influence of the outliers on RKLD loss is much lower than that on the $\ell_2$ and Poisson loss.  Moreover, as observed empirically, unlike the $\ell_2$ and Poisson loss functions, RKLD-based algorithms do not require an adaptive step-size to converge in the gradient descent iterations while requiring fewer iterations than the ones in \cite{candes2015phase, zhang2016reshaped}, and \cite{zhang2018median}.
Unlike the $\ell_2$ and Poisson loss \cite{banerjee2005clustering}, the RKLD loss does not correspond to an underlying assumption on the statistical noise distribution. 
This provides a desirable flexibility for inference using imperfect data, for instance, those due to dynamic imaging environments, or due to interference relevant to applications in wave-based imaging. The RKLD minimization-based approach, therefore, can fit data around outliers and potential model mismatches due to its distribution-agnostic formulation.

Our truncation techniques are inspired from Truncated Wirtinger Flow (TWF)\cite{chen2017solving} and the Median Truncated Wirtinger Flow (median-TWF) algorithms\cite{zhang2018median} which utilize the Poisson-loss function. 
The TWF algorithm \cite{chen2017solving} uses a sample mean-based truncation mechanism. However, in the presence of strong {additive} outliers, it fails to recover the phase information due to high magnitude samples. 
In \cite{zhang2018median}, the properties of median estimator are exploited to develop the median-TWF and median-RWF algorithms using the Poisson and the reshaped $\ell_2$-loss functions, respectively. 
These algorithms demonstrate improved robustness against {additive} noise and sparse outliers than those observed within truncated $\ell_2$-loss minimization \cite{chen2017solving,wang2017solving}, at the cost of near-optimal sample complexity. 
Our numerical study using both synthetic and real data demonstrates that the RKLD-WF and RKLD-MTWF algorithms have superior sample efficiency to those in \cite{candes2015phase} and \cite{zhang2018median}. In \cite{chen2017robust}, the Robust Wirtinger Flow (Robust-WF) algorithm is developed. This algorithm avoids truncation by applying a hard-thresholding technique; however, it requires a priori knowledge of the fraction of corrupted measurements. Unlike Robust-WF, our approach does not require such a priori information.

\subsection{Organization of the Paper}
The rest of the paper is organized as follows. Section \ref{sec:problem_form} describes the problem statement and the formulation of the RKLD-based loss function, followed by a robustness analysis for the RKLD and $\ell_2$ loss functions in Section \ref{sec:robustness}. Section \ref{sec:init} presents the RKLD-based initialization technique. 
The three algorithms, RKLD-WF, RKLD-MTWF, and RKLD-GTWF, are described in Section \ref{sec:algorithms}. Finally, extensive numerical results and comparison of RKLD with the state-of-the-art, using the synthetic and real data, are provided in Section \ref{sec:numerical_experiment} and \ref{sec:real_data}, respectively.


\section{Problem Formulation} \label{sec:problem_form}
 \subsection{The Measurement and Outlier Model}
In practice, the measurements typically contain additive noise and arbitrary outliers \cite{zhang2018median, weller2015undersampled}. Therefore, we express the measurement model in the following vector form:
\begin{equation}\label{eq:phalessvec}
\y = |\mathbf{A}\mathbf{x}|^2+\w + \bfeta,
\end{equation}
where $\y = \left [y_1, \cdots, y_M \right ]^\mathsf{T} \in \Real^M$ is corrupted data, $\mathbf{A} = \left[\a_1^*,\cdots, \a_M^*\right] \in \mathbb{C}^{M\times N}$ is the sampling matrix where $(.)^*$ denotes the conjugate transpose of the measurement vector, $\a_m$. Here, $\w \in \Real^M$ denotes the noise vector that, we assume, is dense and bounded in magnitude so that $\|\w\|_{\infty} \leq \sigma \|\x\|_2^2$ for some constant $\sigma >0$. 
Here, $\bfeta \in \Real_{+}^M$ denotes the outlier vector which is modeled as sparse. We assume that the non-zero outlier elements appear as multiple of the measurement values. Hence, we model the $m$-th outlier element as $\eta_m = \theta_m y_m$, where $\theta_m$ is the non-negative multiplicative factor, typically $\theta_m > 0$. Additionally, we assume that $\theta_m$'s are independently and identically drawn from the following probability distribution:
\begin{equation}\label{eq:outlier_pdf}
   f(\theta_m) = \begin{cases}
    \frac{\rho}{\theta_{max}-\theta_{min}}, & \theta_{min} \leq \theta_m \leq \theta_{max}\\
    1-\rho, & \theta_m = 0\\
    0, & \text{otherwise}
    \end{cases}
\end{equation}
where, $\rho$ is the fraction of the measurements corrupted by outliers and $0 < \theta_{min} \leq \theta_{max}$ are the minimum and maximum positive values, respectively, that $\theta_m$'s can acquire. 

Phase retrieval is an ill-posed problem since multiple signals can result in the same magnitude measurements. Hence, we can recover the unknown only up to a global phase factor.

\subsection{The RKLD-Based Loss Function}

KLD, on the $M$-standard simplex, is a measure of dissimilarity between two probability distributions. More generally, beyond $M$-simplex, KLD measures the dissimilarity between two vectors, $\p, \q \in \Real_+^M$ as follows\cite{yonel2020spectral}: \begin{equation}
\label{eq:KLdiv1}
    D_{KL}(\q,\p) = \sum_{m=1}^{M} q_m\log \frac{q_m}{p_m} - \sum_{m=1}^{M}(q_m - p_m),
\end{equation}
where $p_m$ and $q_m$ are the $m$-th elements of the vectors $\p$ and $\q$, respectively.
KLD is noncommutative in its input arguments, which implies that FKLD, denoted by $D_{KL}(\p,\q)$, and RKLD, denoted by $D_{KL}(\q,\p)$, are not equal.

In this paper, we derive three algorithms and a spectral initialization scheme based on RKLD. Therefore, substituting $\q$ and $\p$ with $|\bA\z|^2$ and $\y$, respectively, in (\ref{eq:KLdiv1}), we define the mismatch between the estimates and the measurements as follows:
\begin{equation}
\begin{split}\label{eq:KL_div}
    \J_{RKLD}(|\mathbf{A}\z|^2,\y)
    &= \sum_{m=1}^{M} |\a_m^*\z|^2\log \frac{|\a_m^*\z|^2}{y_m} - \sum_{m=1}^{M}(|\a_m^*\z|^2 - y_m).
\end{split}
\end{equation} 
 
The RKLD defined in \eqref{eq:KL_div}, is a non-convex function which is nonnegative everywhere and has global minima when the estimate is equal to the true solution. 
\begin{table}[t!]
\centering
\caption{The $\ell_2$ and RKLD loss, with their first derivatives and expected first derivatives. The losses are computed over a single corrupted measurement for a scalar unknown, $x$. The expected derivatives are computed with respect to the real Gaussian measurement map, $a \sim \N(0,1)$.}

\begin{tabular}{|cccc|}
\hline
\multicolumn{4}{|c|}{Measurement model, $y = (1+\theta)|ax|^2 $}                                                         \\ \hline
\multicolumn{1}{|c|}{\begin{tabular}[c]{@{}c@{}}Loss functions\end{tabular}} & \multicolumn{1}{c|}{\begin{tabular}[c]{@{}c@{}}Loss functions, $\J$\end{tabular}} & \multicolumn{1}{c|}{$\partial_{\theta}\J$}                          & $\mathbb{E}[\partial_{\theta}\J]$                                                                                                                                                                 \\ \hline
\multicolumn{1}{|c|}{$\ell_2$}                                                  & \multicolumn{1}{c|}{$\theta^2|ax|^4$}                                                & \multicolumn{1}{c|}{$2\theta|ax|^4$}                              & $6\theta$                                                                                                                                                                                       \\ \hline
\multicolumn{1}{|c|}{$RKLD$}                                                    & \multicolumn{1}{c|}{$|ax|^2\left(\theta -\log(1+\theta)\right)$}                     & \multicolumn{1}{c|}{$|ax|^2\left(\frac{\theta}{\theta+1}\right)$} & $\frac{\theta}{\theta+1}$                                      \\ \hline
\multicolumn{1}{|c|}{$Poisson$}                                                    & \multicolumn{1}{l|}{\begin{tabular}[c]{@{}l@{}}$(1+\theta)|ax|^2 $$\log(1+\theta)-\theta|ax|^2$\end{tabular}}                      & \multicolumn{1}{c|}{$|ax|^2\log(1+\theta)$} & $\log(1+\theta)$                                                                                                                                 \\ \hline
\end{tabular}
\label{tab:single_meas}
\end{table}

When the measurements are Poisson distributed, the corresponding negative log-likelihood function is equivalent to the FKLD\cite{chen2017solving, zhang2018median}. In FKLD, the log-mismatch term, $\log \left(\frac{|\a_m^*\z|^2}{y_m}\right)$ is weighted with the corresponding actual measurement, $y_m$ whereas, in RKLD, we use the estimated measurements, $|\a_m^*\z|^2$ as the weighting factors, which are refined during each iteration. Thus, the optimization landscape provided by RKLD is significantly different from that of FKLD. 

\section{Robustness of the RKLD Loss}\label{sec:robustness}
\subsection{With Respect to Outliers}
The key features of our interest in RKLD stem from the logarithmic processing of the underlying measurements, $\{y_m \}_{m =1}^M$. Firstly, the logarithmic processing naturally suppresses outliers in the data by shrinking the dynamic range of measurements. 
The logarithmic processing prevents strong outliers from heavily influencing the gradient computation by reducing the multiplicative factors to additive residuals. To demonstrate the superiority of RKLD-based optimization over that of the $\ell_2$ and Poisson loss functions in suppressing the effect of outliers, we study the influence curves\cite{hampel1974influence}, i.e., the derivatives of the loss functions with respect to the outlier. 

Since each sample contributes to the loss function separately, we consider a a single corrupted measurement, $y = |ax|^2+\eta$, where $\eta = \theta y$ with $\theta \sim f(\theta)$. 
Here, we assume that the measurement map, $a$ and the ground truth, $x$ are one dimensional where $a \sim \mathcal{N}(0,1)$ and $|x|=1$. 

At a global optimum, the loss values are equal to 0. 
However, the corruption of the measurements by outliers causes the loss functions to deviate from their ideal values. 
To assess the impact of the outliers in the deviation from the ideal loss functions, we study their partial derivatives with respect to the outlier magnitude parameter, $\theta$. 
It is desirable to have this partial derivative to be bounded to control the deviation in case of large $\theta$ values. 
Table 1 provides the loss functions, their derivatives with respect to $\theta$, and the expectation of the derivatives with respect to the sampling model, $a$. Here, it suffices for $a$ to have finite second and fourth order statistics, which is the case for the Gaussian and coded diffraction measurement maps under study.

\begin{figure}
    \centering
    \includegraphics[width = 7cm]{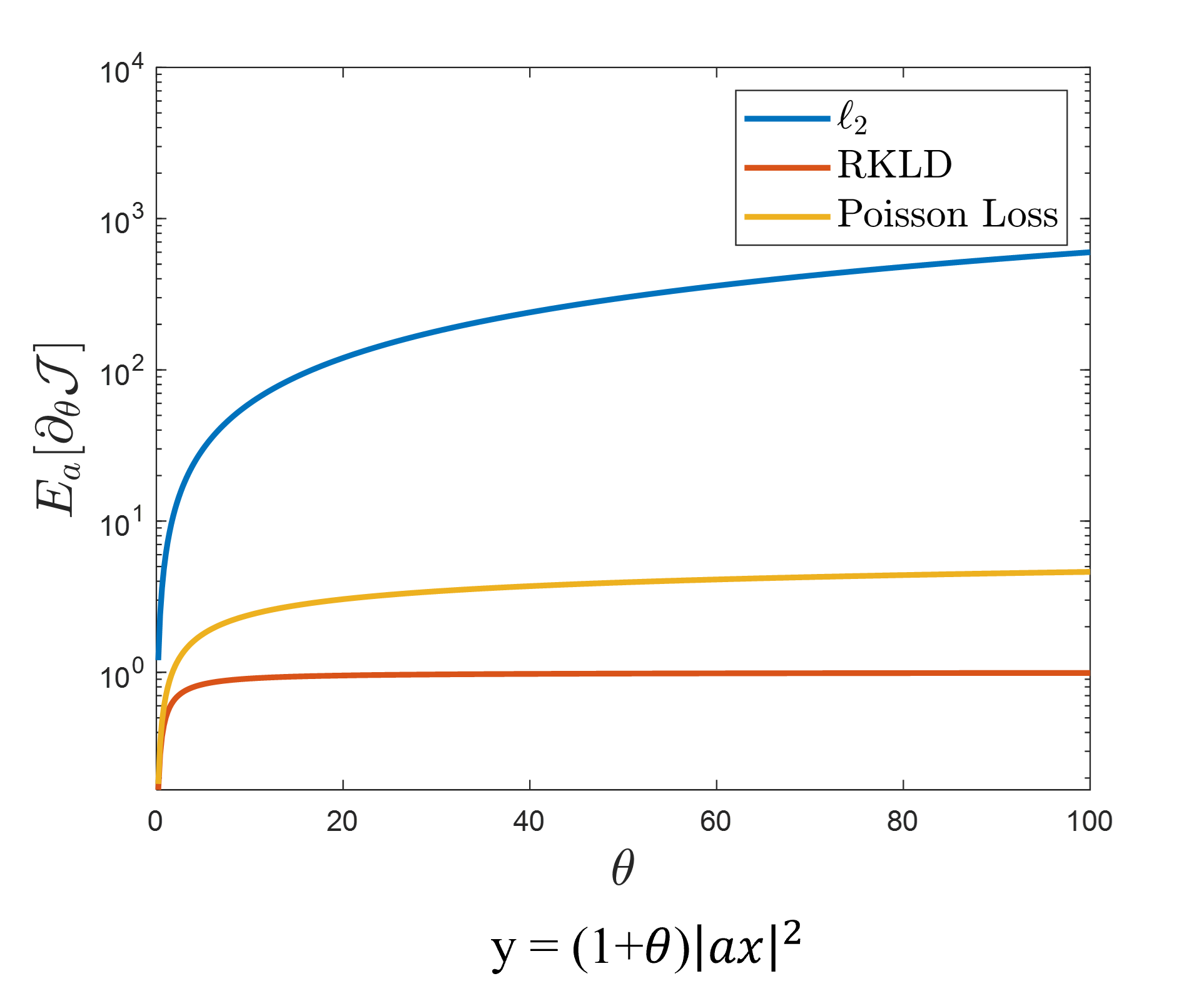}
    \caption{The log of expected gradients, $\mathbb{E}_a[\partial_\theta \J]$ with respect to the outlier magnitude, $\theta$. The $\ell_2$ gradient increases linearly with $\theta$ whereas the RKLD gradient levels off at $1$ for large $\theta$.}
    \label{fig:gradient_theta}
\end{figure}

Figure \ref{fig:gradient_theta} shows the expected derivatives of each loss function with respect to $\theta$ in log-scale. 
Clearly, the unbounded nature of the expected derivative of the $\ell_2$ loss function, with $\mathbb{E}_a[\partial_{\theta}\J_{\ell_2}] = 6 \theta$, shows that $\mathcal{J}_{\ell_2}$ is highly influenced by the outliers. For the Poisson loss function also, $\mathbb{E}_a[\partial_{\theta}\J_{Poisson}] = \log(1+\theta)$ indicates that as $\theta$ increases, the impact of outliers increases.
On the other hand, the RKLD loss function has a bounded expected derivative with respect to $\theta$, with in fact, $\mathbb{E}_a[\partial_{\theta}\J_{RKLD}] \rightarrow 1$ as $\theta \rightarrow \infty$.

Similarly, for $\eta = \theta x^2$, $\mathbb{E}[\partial_\theta \J_{\ell_2}] = 2\theta$ and $\mathbb{E}[\partial_\theta \J_{Poisson}] = \int \log\frac{u^2+\theta}{u^2}e^{-u^2/2}du$ are unbounded as $\theta \rightarrow \infty$, whereas $\mathbb{E}[\partial_\theta \J_{RKLD}] = \sqrt{\frac{\pi\theta}{2}}e^{\theta/2}\text{erfc}\left(\sqrt{\frac{\theta}{2}}\right)$ gradually approaches $1$ as $\theta \rightarrow \infty$.
Hence, the influence of outliers on the RKLD loss function is limited around the true solution. 
Thus RKLD shows more robustness against large outliers than $\ell_2$.  

\subsection{With Respect to Noise}\label{sec:robust-noise}

Another key property is the fact that logarithmic processing promotes orthogonality. 
This is a consequence of the mode-seeking property in probability density estimation, since RKLD is biased to fit the support of the target distribution, instead of an average fit over the event set\cite{wainwright2008graphical}. 
This can be observed as $y_m \rightarrow 0$, that is, for noise only measurements, at any estimate that does not satisfy $| \a_m^* \z |^2 \rightarrow 0$ the loss function $\J_{RKLD}$ diverges. Essentially, $y_m = 0$ inserts a hard-constraint for the estimates, as the set of sampling vectors $\{ \a_m \}_{I(y=0)}$ provide an orthogonal subspace to the solution, where $I(y=0)$ denotes the index-set of the measurements that are equal to $0$. 



Note that this orthogonal subspace projection is an implicit property of $\J_{RKLD}$, and it is approximately effective for small measurements in the sense that, as $y_m \rightarrow 0$, $\J_{RKLD} \rightarrow \infty$. We can express the loss function in an alternative perspective to demonstrate its orthogonality promoting property as follows:
\begin{equation}\label{eq:ortho}
\J_{RKLD}(\z) = 
\tilde{\J}_{RKLD}(\z) + i\left( \z \perp \{ \a_m \}_{m \in I(y=0)}\right), \end{equation}
where $i(\cdot)$ is the indicator function that assigns $\infty$ to estimates that do not reside in the constraint set (the loss is identical if the index set consists of $y_m = 0$ precisely, which is analogous to a noise subspace) and $\tilde{\J_{RKLD}}(\z)$ takes the following form:
\begin{equation}
\tilde{\J}_{RKLD}(\z) = \sum_{m \not \in I(y=0)}^M \left( |\a_m^*\z|^2\log \frac{|\a_m^*\z|^2}{y_m} - |\a_m^*\z|^2 \right).
\end{equation}

This constraint on the search space provides further robustness against outliers present in the data, since the inherent projection operation can mitigate the influence of the outliers on the iterative estimates.


In the following sections, to simplify our notations, we drop the subscripts from $\J$ and use it to denote the RKLD loss function, unless stated otherwise.


\section{Initialization Based on RKLD}\label{sec:init}

Accuracy of the initial estimates provided by spectral methods is known to facilitate the exact recovery guarantees for non-convex iterative approaches\cite{yonel2020deterministic,candes2015phase,netrapalli2013phase,chen2017solving}, under the Gaussian model and coded diffraction patterns. 
The classical spectral method involves setting the leading eigenvector of the following empirical scatter matrix:
\begin{equation}\label{eq:spectral}
\mathbf{Y} = \frac{1}{{M}}\sum_{m = 1}^{{M}} y_{m} \mathbf{a}_m \mathbf{a}_m^*
\end{equation}
which is coupled with a norm estimate of the signal to initialize the algorithm. 

In our work, we obtain an initial estimate of the unknown signal using the spectral estimation framework that we have recently introduced in \cite{yonel2020spectral}. 
In \cite{yonel2020spectral}, we consider the classical formulation of the spectral method as an approximate minimization of the $\ell_2$-loss function, which is equivalent to maximizing the correlation between the lifted model and the spectral matrix, $\mathbf{Y}$.
The accuracy of the spectral methods are interpreted under a certain restricted isometry-type property related to the sufficient conditions of the WF-based exact phase retrieval algorithm\cite{yonel2020deterministic}.
This observation is then generalized to Bregman divergences, yielding the sample processing functions $h$ that can be obtained by solving the following optimization problems: 
\begin{align}
(P1)&: \underset{\q \in X \subset \mathit{S}}{\text{minimize}} \ d(\q, \p) \approx \ - \langle \q, h(\p, \hat{\q}) \rangle \label{eq:dmineq1}\\
(P2)  &: \underset{\| \v \| = 1}{\text{maximize}} \ \v^H \left[\sum_{m = 1}^M h(\p, \hat{\q})_m \ \a_m \a_m^* \right] \v, \label{eq:dmineq2}
\end{align}
Here, $P1$ represents the approximate minimization of the Bregman divergence over $\q = | \langle \a_m , \v \rangle|^2$, where $\v$ is the spectral search variable. This approximation is facilitated by a sample processing function, $h$, which is designed to introduce minimal distortion to the original objective using the Bregman representation property \cite{banerjee2005optimality}, denoted by $\hat{\q}$. 
Hence, $P2$ represents an alternative to \eqref{eq:spectral}, such that the Bregman divergence minimization is approximated by a tractable spectral method. 
The two problems are complementary in the sense that the maximization in $P2$ sufficiently approximates the minimization of the loss in $P1$ through the derivation of $h$. 

To construct the spectral estimate for the KL-divergence loss, we consider:
\begin{equation}\label{eq:KLspec}
D_{KL}(\q, \p) = \sum_{m = 1}^M q_m \log \frac{q_m}{p_m},
\end{equation}
constrained on the M-simplex. 
We set $\p = \y / \| \y \|_1$. Then the RKLD minimizing spectral method is formulated under minimum distortion as follows:
\begin{equation}
   D_{KL}(\q, \p) \approx - \sum_{m = 1} q_m \log \left( \frac{y_m / \| \y \|_1}{\| \a_m \|_2^2 / \sum_{i = 1}^M \| \a_i \|_2^2} \right)\label{eq:KLaprx}
\end{equation} 
\vspace{.1cm}
\begin{equation}
   \z_0 := \underset{\| \v \| = 1}{\text{argmax}} \ \v^H \left[\sum_{m = 1}^M \log \left( \frac{y_m \sum_{i = 1}^M \| \a_i \|_2^2}{\| \a_m \|_2^2 \| \y \|_1}  \right) \a_m \a_m^* \right] \v. \label{eq:KLfinal}
\end{equation}

Algorithm 1 summarizes the pseudo-code to obtain the RKLD based spectral initialization.
\begin{algorithm}[!h]
\caption{RKLD Initialization}
\label{alg:init}
\begin{algorithmic}

\STATE{\textbf{Input:}} The measurements, $\{\y\}_{m=1}^{M}$, the sampling vectors, $\{\a_m\}_{m=1}^{M}$
\STATE{\textbf{Step 1.}} Compute the sample processing function $$h_m = \log \left( \frac{y_m / \| \y \|_1}{\| \a_m \|_2^2 / \sum_{i = 1}^M \| \a_i \|_2^2}\right)$$
\STATE{\textbf{Step 2.}} $\z_0 = \underset{\| \v \|_2 = 1}{\text{argmax}} \ \v^* \left[ \sum_{m = 1}^M h_m \a_m \a_m^* \right]\v$
\STATE{\textbf{Output:}} The leading eigenvector, $\z_0$
\end{algorithmic}
\end{algorithm}

\section{Phase Retrieval Based on RKLD $\&$ the WF Framework}\label{sec:algorithms}
In this section, we introduce and describe the corresponding WF algorithms. 

\subsection{RKLD-WF Algorithm}
We formulate the following loss function, $\J(\z)$, for the RKLD-WF algorithm:

\begin{equation}
\begin{split}\label{eq:objfunkl1}
 \J(\z) &\coloneqq \frac{1}{2}
\sum_{m = 1}^M \left(|\a_m^*\z|^2\log \frac{|\a_m^*\z|^2}{y_m} -|\a_m^*\z|^2\right). 
\end{split}
\end{equation}

Since $\J(\z)$ is a real-valued function of a complex variable, it is non-holomorphic and not complex differentiable \cite{hunger2007introduction}. Hence, we compute the Wirtinger derivative of $\J(\z)$ \cite{candes2015phase},

\begin{equation}\label{eq:defgrad}
    \nabla \J(\z) = \left(\frac{\partial \J}{\partial \z}\right)^*,
\end{equation}
which results in, 
\begin{equation}\label{eq:grad_kl}
\begin{split}
    \nabla \J(\z)
    & = \sum_{m=1}^{M} \left[\log\left(|\a_m^* \z|^2\right)- \log \left(y_m\right)\right] \left(\a_m^*\z\right)\a_m.
\end{split}
\end{equation}

\eqref{eq:grad_kl} can be compactly expressed in terms of matrix-vector products as follows:
\begin{equation}\label{eq:gradkl_A}
    \nabla \J(\z) = \bA^*\left[\bA\z \odot \left(\log\left(|\bA\z|^2\right) - \log\left(\y\right) \right)\right],
\end{equation}
where, $\bA^*$ denotes the conjugate transpose of $\bA$, and $\odot$ denotes the element-wise multiplication of two vectors. 

Due to the logarithmic processing of the quantities in (\ref{eq:gradkl_A}), the Wirtinger gradient is not defined when one or more components of $\bA\z$ are zero (or close to zero). This motivates the use of a regularized loss function. Therefore, instead of (\ref{eq:objfunkl1}), we introduce the following alternative loss function:
\begin{equation}\label{eq:reg_kl}
   \J_{\lambda}(\z) := \sum_{m = 1}^M (|\a_m^*\z|^2+\lambda)\log \frac{|\a_m^*\z|^2+\lambda}{y_m+\lambda}
    - \sum_{m=1}^{M}(|\a_m^*\z|^2+\lambda),
\end{equation}
where $0< \lambda \ll 1$. The corresponding Wirtinger gradient expression is given as follows:
\begin{equation}\label{eq:gradkl_delta}
 \nabla \J_\lambda(\z) = \bA^* [\bA\z \odot (\log(|\bA\z|^2+\lambda\boldsymbol{1})-\log(\y+\lambda\boldsymbol{1}))],
 \end{equation}
where $\boldsymbol{1}$ is an all-ones vector of length $M$.
 
For a fixed step-size, $\mu$, we write the update equation for the gradient descent iterations as follows: 
\begin{equation}\label{eq:GD}
    \z_{k+1} = \z_{k} - \mu \nabla \J_{\lambda}(\z_{k}).
\end{equation}

Algorithm \ref{alg:WF-KLD} summarizes the pseudo-code for the regularized RKLD-WF algorithm.

\begin{algorithm}
\caption{Regularized RKLD-WF}
\label{alg:WF-KLD}
\begin{algorithmic}
\STATE{\textbf{Input:}} $\{\y\}_{m=1}^{M}, \{\a_m\}_{m=1}^{M}$, Step size, $\mu$, regularization parameter, $0< \lambda \ll 1$.
\STATE{\textbf{Step 0.}} Initialize $\z_0$  following Algorithm \ref{alg:init}. 
\FOR{$k=0$ to $(K-1)$}
\STATE{\textbf{Step 1.}} Compute $\nabla \J_{\lambda}(\z^{(t)})$ as in (\ref{eq:gradkl_delta})
\STATE{\textbf{Step 2.}} $\z_{k+1} = \z_{k} - \mu\nabla \J_{\lambda}(\z_{k})$
\ENDFOR
\STATE{\textbf{Output:} $\z_K$}
\end{algorithmic}
\end{algorithm}

\subsection{Residual Truncation with RKLD}\label{sec:residual_trunc}
To further improve the performance of the RKLD-WF algorithm against outliers, we propose to control the contribution of the outlier-corrupted measurements to the total loss by truncating the sampling vectors. This leads to the formulation of the following truncated RKLD-based loss function:
\begin{equation}
    \J^{t}(\z) 
    \coloneqq \sum_{m = 1}^M \{|\a_m^*\z|^2\log \frac{|\a_m^*\z|^2}{y_m} 
    -|\a_m^*\z|^2 \} \mathds{I}_{\boldsymbol{\Tau}}(m),
\end{equation}
where $\mathds{I_{\boldsymbol{\Tau}}(m)}$ denotes the $1/0$ indicator function for the index set $\boldsymbol{\Tau} \subset \{1, 2, \cdots, M \}$ determined by the truncation scheme. Thus, instead of using all $M$ samples, we use an adaptive collection of suitable $M_k^\prime \leq M$ sampling vectors, $\{\a_m\}_{m \in \boldsymbol{\Tau}}$ to compute the total loss at each iteration.


For the construction of the index set, $\Tau$, the most straight forward approaches are based on the \emph{absolute residuals}, i.e., the model mismatch at a given estimate $\z$, $\left\{| y_m - | \a_m^* \z |^2 |\right\}_{m=1}^M$. 
The truncation scheme can be designed with respect to the mean or the median of the absolute residuals. 
These mean and median-based truncation mechanisms are employed in \cite{chen2017solving}, and \cite{zhang2018median}, respectively, where the total loss is modified by carefully choosing from the sampling vectors so that the selected sampling vectors satisfy certain statistical criteria conditioned upon the residuals. In the TWF algorithm \cite{chen2017solving}, the truncation scheme forms the following index sets, $\Tau_1$ and $\Tau_2$, which are adaptively modified at each iteration:
\begin{gather}
    \Tau_1 := \left\{m: \gamma^{lb}\leq\frac{|\a_m^*\z|}{\|\z\|_2}\leq\gamma^{ub}\right\}, \label{eq:ind1}\\
    \Tau_2 \coloneqq \left\{ m : \left| \ y_m -|\a_m^*\z|^2 \ \right|\leq \gamma^e K_l\right\},\label{eq:ind2}\\
    K_l := \text{mean}(\left\{\left| \ y_m - |\a_m^*\z|^2 \ \right| \right\}_{m=1}^{M}), \label{eq:mean}
\end{gather}
where $\gamma^{lb}$, $\gamma^{ub}$, and $\gamma^e$ are the predetermined threshold parameters\cite{chen2017solving}, and $K_l$ denotes the sample mean of the residuals.

 The sample mean, $K_l$ serves as the $\ell_2$-loss minimizer of the residuals. 
Minimizing the $\ell_2$-loss of the residuals comes from the assumption that the residuals identically and independently follow a Gaussian distribution. 
However, the sample median is a more robust estimate with respect to outliers than the sample mean.  
Hence, we assume the residuals follow a Laplace distribution and the negative log-likelihood of this distribution yields an $\ell_1$ data fit term based on the residuals. Finally, by minimizing the $\ell_1$ loss, we get the following estimate of the residuals, which is the sample median: 
\begin{gather}
    K_{med} := \text{median}(\left\{\left| \ y_m - |\a_m^*\z|^2 \ \right| \right\}_{m=1}^{M}),
\end{gather}
Therefore, we use $K_{med}$ in our truncation scheme as a more robust estimate of the residuals against the outliers. Additionally, from the RKLD gradient, $\{\log(|\a_m^*\z|^2)-\log(y)\}(\a_m^*\z)\a_m$, we see that $|\a_m^*\z|$ being small does not make the gradient large, unlike the Poisson loss function. Hence, we discard the lower bound in \eqref{eq:ind1} and modify \eqref{eq:ind1} for RKLD. Finally, we rewrite the truncation schemes for our RKLD-based median truncated WF (RKLD-MTWF) algorithm as following:
\begin{gather}
    \Tau_1 := \left\{m: \frac{|\a_m^*\z|}{\|\z\|_2}\leq\gamma^{ub}\right\}, \label{eq:ind11}\\
    \Tau_2 \coloneqq \left\{ m : \left| \ y_m -|\a_m^*\z|^2 \ \right|\leq \gamma^e K_{med}\right\},\label{eq:ind21}\\
    K_{med} := \text{median}(\left\{\left| \ y_m - |\a_m^*\z|^2 \ \right| \right\}_{m=1}^{M}), \label{eq:mean1}
\end{gather}
where the thresholding parameters $\gamma^{ub}$, $\gamma^{lb}$ and $\gamma^e$ are manually tuned.

We summarize the pseudo-code for RKLD-MTWF in Algorithm \ref{alg:KLD-MTWF}. 
\begin{algorithm}
\caption{RKLD-MTWF}
\label{alg:KLD-MTWF}
\begin{algorithmic}
\STATE{\textbf{Input:}} $\{\y\}_{m=1}^{M}, \{\a_m\}_{m=1}^{M}$, step size, $\mu_{mtr}$, truncation parameters, $\gamma^{ub}$, and $\gamma^{e}$.
\STATE{\textbf{Step 0.}} Initialize $\z_0$ following Algorithm \ref{alg:init}. 
\FOR{$k=0$ to $(K-1)$}
\STATE{\textbf{Step 1.}} Compute, $K_l := \text{median}(\left\{|y_m - |\a_m^*\z_k|^2\right\}_{m=1}^{M})$

\STATE{\textbf{Step 2.}} Compute, $\mathbf{I}_k = [\mathds{I}_{\Tau_1\cap\Tau_2}(1), \cdots, \mathds{I}_{\Tau_1\cap\Tau_2}(M)]^T$
where,\begin{equation}\label{eq:ind111}\nonumber
    \Tau_1 \coloneqq \left\{ m : \frac{|\a_m^*\z_k|}{\|\z_k\|_2} \leq \gamma^{ub}\right\}, \quad
    \Tau_2 \coloneqq \left\{ m : \left|y_m -|\a_m^*\z_k|^2\right|\leq \gamma^e K_l\right\}
\end{equation}
\STATE{\textbf{Step 3.}} Compute, $\bA_k^t = \bA \odot \mathbf{I}_k$ and $\y_k^t = \y \odot \mathbf{I}_k$
\STATE{\textbf{Step 4.}} Compute,
$\nabla \J^{t}(\z_k) = {\bA^t}^*_{k}\left[\bA^t_{k}\z_{k} ~\odot \left(\log|\bA^t_{k}\z_{k}|^2-\log\y^t_{k}\right) \right]$

\STATE{\textbf{Step 5.}} Update, $\z_{k+1} = \z_{k} - \mu_{mtr}\nabla \J^{t}(\z_k)$
\ENDFOR
\STATE{\textbf{Output:} $\z_K$}
\end{algorithmic}
\end{algorithm}


\subsection{Gradient Truncation with RKLD}

Residual-based truncation schemes are natural for the general class of bowl-shaped loss functions because these functions are defined directly in terms of the residual norms. The residual term, $(y_m - |\a_m^*\z|^2)$ also appears as the weights of the sampling vector, $\a_m$ in the synthesis of the gradient. Since the total gradient is the summation of these weighted sampling vectors, the sampling vectors with large weights dominate the direction of the update. These large weights may appear due to the outliers in the measurements or model mismatch. Hence, it is important to control the weights the sampling vectors via truncation mechanisms.

For $\ell_2$ and Poisson loss functions, the residual truncation immediately translates to the truncation of the weights. However, for RKLD,  the residual is not the weights of the sampling vectors in the gradient. Instead, the weights are the residual of logarithms, $\left[\log\left(|\a_m^* \z|^2\right)- \log \left(y_m\right)\right]$.
Hence, the residual-based truncation does not equivalently translate to controlling the weights of sampling vectors within the gradient of RKLD. 
To address this fundamental difference between the RKLD and the other loss functions, we take inspiration from  \cite{chen2017solving}, and introduce a gradient-based truncation scheme which uses the median of $\left[\log\left(y_m\right)- \log \left(|\a_m^* \z|^2\right)\right]$ as the truncation criteria.
Thus, for the gradient truncation, we develop the following one-sided truncation scheme:
\begin{equation}\label{eq:single-sided}
    \Tau_3 \coloneqq \left\{ m ~|~ \left(\log \left(y_m\right)-\log\left(|\a_m^* \z_k|^2\right)\right)\leq \gamma^h \mathcal{D}\right\},
\end{equation}
where,
\begin{equation}
    \mathcal{D} \coloneqq \text{median}\left\{ \log(y_m) - \log(|\a_m^*\z_k|^2) \right\}_{m=1}^{M},
\end{equation}
with $\gamma^h$ denoting a constant threshold parameter determined by hand-tuning and $\mathcal{D}$ denoting the median of the \emph{residual of logarithms}. 
The truncation scheme in (\ref{eq:single-sided}) is designed such that the sampling vectors, demonstrating the orthogonality promoting properties of RKLD, are preserved. 

The conventional two-sided truncation is based on the statistics of the absolute residuals of logarithms, $\{| \log |\a_m^*\z|^2  - \log y_m |\}_{m=1}^M$.
Notably, the absolute residual corresponds to:
\begin{enumerate}
\item $\log |\a_m^*\z|^2 - \log y_m > 0$ if $|\a_m^*\z|^2 > y_m$, or
 \item  $\log y_m - \log |\a_m^*\z|^2 > 0$ if $y_m > |\a_m^*\z|^2$.
\end{enumerate}

When $y_m = 0$ and the model prediction is $| \a_m^* \z |^2 \neq 0$, we get $|\log | \mathbf{a}_m^* \z |^2 - \log y_m | \rightarrow \infty$. Then, the ideal action is to enforce such orthogonality by using the sampling vectors in the orthogonal subspace, $\{\a_m\}_{I(y=0)}$ as discussed in Section \ref{sec:robust-noise}. 
However, in this case, $|\log | \mathbf{a}_m^* \z |^2 - \log y_m |$ being large does not necessarily indicate an outlier in the data; on the contrary, it potentially contains key information about the unknown which we wish to leverage by the use of RKLD. Thus, a two-sided truncation criterion removes orthogonality information from the synthesis for $y_m \rightarrow 0$.

Clearly, the samples, $y_m = 0$ are included in the first case ($|\a_m^*\z|^2 \geq 0$ by definition) which we aim to preserve even after truncation. 
On the other hand, when the data is contaminated with outliers, the second case occurs where we observe the detrimental effect in generating estimates. 
Therefore, we can employ a \emph{single-sided} scheme, where only the residual of logarithms in the form of case (2) are truncated, thus preserve the noise subspace projections. This truncation mechanism leads to the following gradient expression:
\begin{equation}
   \nabla \J^{t}(\z) = (\bA^{t})^*[\bA^{t}\z ~\odot  (\log(|\bA^{t}\z|^2)-\log(\y^{t})) ],
\end{equation}
where $\bA^{t}$ and $\y^{t}$ are the truncated measurement matrix and the measurements, respectively, containing components corresponding to the indices in $\Tau_3$ which gets updated at each iteration.

The pseudo-code for the RKLD-based gradient truncated WF (RKLD-GTWF) algorithm is summarized in Algorithm \ref{alg:KLD-GMTWF}.
 \begin{algorithm}
 \caption{RKLD-GTWF with one-sided truncation}
 \label{alg:KLD-GMTWF}
 \begin{algorithmic}
 \STATE{\textbf{Input:}} $\{\y\}_{m=1}^{M}, \{\a_m\}_{m=1}^{M}$, step size, $\mu_{gtr}$, truncation parameter, $\gamma^{h}$.
 \STATE{\textbf{Step 0.}} Initialize, $\z_0$ following Algorithm \ref{alg:init}. 
 \FOR{$k=0$ to $(K-1)$}
 \STATE{\textbf{Step 1.}} Compute,
$\mathcal{D} \coloneqq \text{median}\left\{ \log(y_m) - \log(|\a_m^*\z_k|^2) \right\}_{m=1}^{M}$.
 
 \STATE{\textbf{Step 2.}} Compute, $\mathbf{I}_k = [\mathds{I}_{\Tau_3}(1), \cdots, \mathds{I}_{\Tau_3}(M)]^T$
 where,
 \begin{equation}\nonumber
     \Tau_3\coloneqq \left\{ m ~|~ \left(\log \left(y_m\right)-\log\left(|\a_m^* \z_k|^2\right)\right)\leq \gamma^h \mathcal{D}\right\}
 \end{equation}
 \STATE{\textbf{Step 3.}} Compute,  $\bA_k^t = \bA \odot \mathbf{I}_k$ and $\y_k^t = \y \odot \mathbf{I}_k$
\STATE{\textbf{Step 4.}} Compute, $\nabla \J^{t}(\z_k) = {\bA^t}^*_{k}\left[\bA^t_{k}\z_{k} ~\odot \left(\log|\bA^t_{k}\z_{k}|^2-\log\y^t_{k}\right) \right]$
 \STATE{\textbf{Step 5.}} Update, $\z_{k+1} = \z_{k} - \mu_{gtr}\nabla \J^{t}(\z_k)$
 \ENDFOR
 \STATE{\textbf{Output:} $\z_K$}
 \end{algorithmic}
\end{algorithm}

\subsection{Computational Complexity}
We now discuss the computational complexity of the RKLD-WF, RKLD-MTWF and RKLD-GTWF algorithms. Gradient computation is the most expensive step in each algorithm, which performs two matrix multiplications, i. e., $\bA^* \v$, and $\v = \bA\z$. Hence, the per-iteration computational complexity of each algorithm is $\mathcal{O}(MN)$, which is equivalent to the per-iteration computational complexity of the state-of-the-art methods \cite{zhang2016reshaped, chen2017solving,wang2017solving, zhang2018median}.  

The initialization procedure entails the computation of the leading eigenvector, which has $\mathcal{O}(N^3)$ complexity, identical to the computational complexity of the classical spectral method.

\begin{figure*}[!ht]
    \centering    \includegraphics[width=\textwidth]{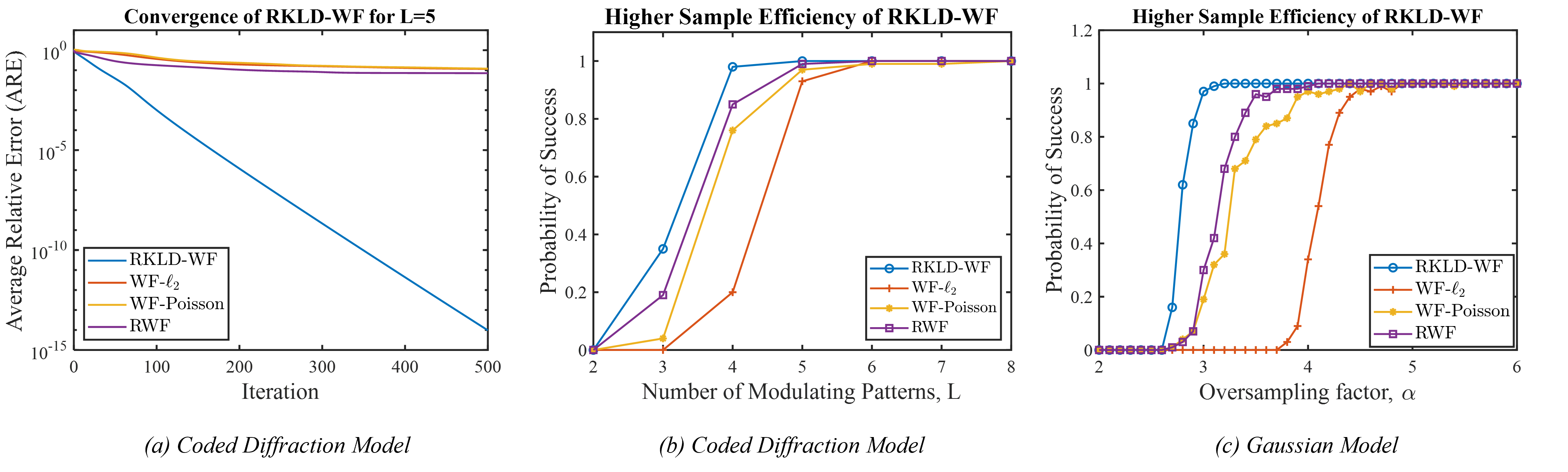}
    \caption{Comparison of the \emph{RKLD-WF} (Algorithm \ref{alg:WF-KLD}) with the standard WF algorithm\cite{candes2015phase} using the $\ell_2$-loss and Poisson-loss functions, and the RWF algorithm \cite{zhang2016reshaped} using the reshaped $\ell_2$-loss function. The results are produced using the noiseless data. \textit{(a)} shows the Average Relative Error (ARE) vs. iteration index for the Coded diffraction model with $N=512$ and the number of modulating pattern $L=5$. \textit{(b)} shows the empirical probability of success w.r.t. the number of modulating pattern ($L$) with $N=256$ for the CDP model. \textit{(c)} shows the empirical probability of success w.r.t. oversampling factors ($\alpha$) with $N=100$ for the Gaussian model. Each result is obtained over 100 Monte-Carlo simulations.}
    \label{fig:wfplot1}
\end{figure*}

\section{Performance Evaluation Using Synthetic Data}\label{sec:numerical_experiment}
We conducted numerical simulations to evaluate the performance of our RKLD-WF, RKLD-MTWF, and RKLD-GTWF algorithms using the CDP and Gaussian forward model. Additionally, we compare the efficiency of these algorithms with their $\ell_2$ and Poisson loss-based counterparts in terms of robustness, convergence speed, and sample efficiency. 



 
\subsection{Performance Evaluation Metrics}\label{sub:perf.eval.metrics}
 
We performed \emph{Monte-Carlo (MC)} simulations for performance evaluation by generating $S$ independent and uncorrelated realizations of the measurement matrix, $\bA$, and corresponding $S$ sets of measurements for a fixed unknown, $\x$. We used the following as a figure of merit, which we refer to as the Average Relative Error (ARE) to evaluate the quality of reconstructed signal:
\begin{equation}\label{eq:avg_err}
    \text{ARE} := \frac{1}{S}\sum_{s=1}^{S}\frac{\text{dist}(\x,\z_K^s)}{\left\|\x\right\|_2},
\end{equation}
where $z^s_K$ denotes the recovered signal using the $s$-th set of measurements after $K$ iterations. Here, $\text{dist}(\x,\z^s_{K})$ is the Euclidean distance between two vectors $\x$ and $\z^s_K$, which is defined as follows:
\begin{equation}\label{eq:dist}
    \text{dist}(\x, \z_K) := \min_{\phi \in [0, 2\pi)} \left\|\x e^{\textit{i}\phi} - \z_K\right\|.
\end{equation} 

To demonstrate the sample efficiency of our algorithms, we used the empirical probability of successful recovery as a figure of merit. We considered the outcome of the $s$-th \emph{MC} simulation  to be successful, if it satisfied $\text{dist}(\x,\z^s_K) < 10^{-5}$. The empirical probability of success (EPS) is defined as follows:

\begin{equation}\label{eq:emps}
    \text{EPS} := \frac{\textit{No. of successful simulations}}{\textit{Total no. of simulations}}.
\end{equation} 

We varied the oversampling factor, $\alpha$, and ran \textit{MC} simulations for each $\alpha$. In addition, we investigated the performance of algorithms at different levels of signal-to-noise ratio (SNR) defined as follows:
\begin{equation}
    \text{SNR} := 10\log_{10}\frac{\text{std}_y}{\text{std}_w},
\end{equation}
where $\text{std}_y$ and $\text{std}_w$ are the standard deviation of the data and the additive noise, respectively. 
\begin{figure}
    \centering   \includegraphics[width=0.4\columnwidth]{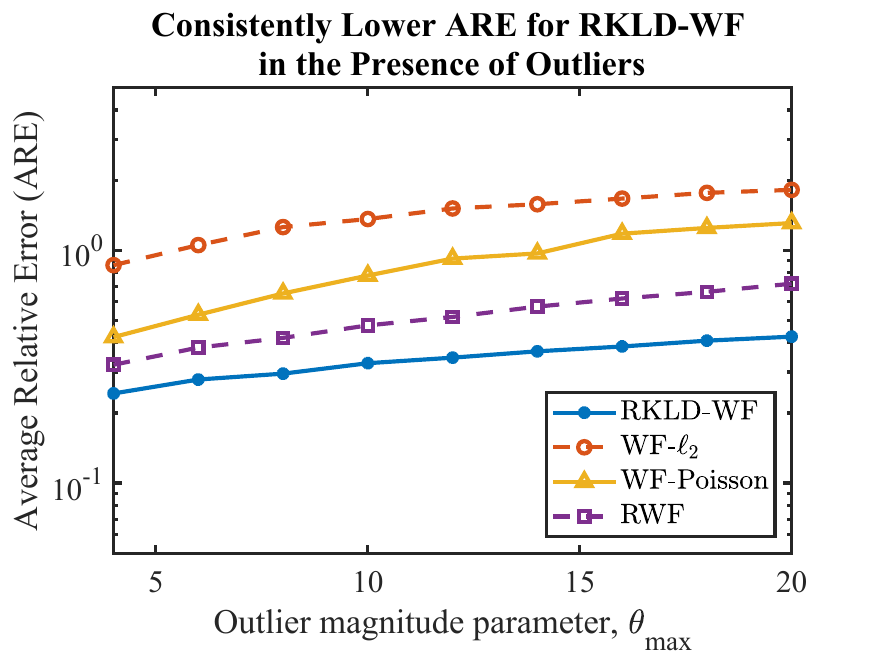}
    \caption{ARE with respect to the outlier magnitude parameter for the RKLD-WF, WF-$\ell_2$, WF-Poisson and RWF algorithms using the CDP model. We set the signal length to $N=128$ and the oversampling factor to $L=8$.}
    \label{fig:robust_WF_l2_kl}
\end{figure}

\begin{figure*}[t]
    \centering
    \includegraphics[width=\textwidth]{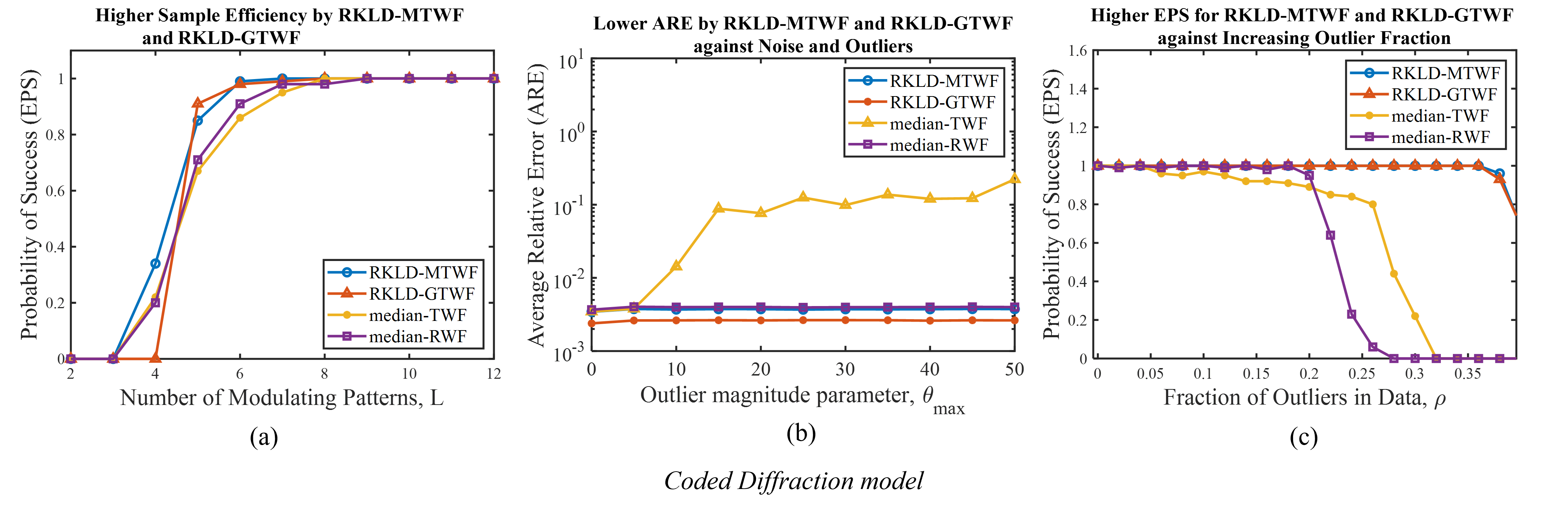}
    \caption{Comparison among the RKLD-MTWF (Algorithm \ref{alg:KLD-MTWF}), RKLD-GTWF (Algorithm \ref{alg:KLD-GMTWF}), median-TWF and median-RWF \cite{zhang2018median} algorithms in terms of sample efficiency and reconstruction accuracy in the presence of outlier and noise. (\emph{a}) EPS with respect to increasing $L$. (\emph{b}) ARE vs. outlier magnitude, $\theta$. (\emph{c}) EPS vs. outlier fraction, $\rho$.}
    \label{fig:mtwfplot1}
\end{figure*}

\subsection{Simulation Set-up and Parameters of Signal and Measurement Models}\label{sub:sim_setup}
\subsubsection{Signal and Measurement Model}
For our simulations, we generated a complex ground truth signal, $\x \sim \mathcal{C}\mathcal{N}(0,\mathbf{I})$ of length $N$. Additionally, owing to the physical significance and prevalence across different imaging modalities, we used a real ground-truth signal, $\x \sim \mathcal{N}(0,\mathbf{I})$.

Our simulation set-up primarily includes the coded diffraction model \cite{candes2015phase2}. The coded diffraction patterns (CDP) are a collection of intensity data, $\y$, which are modeled as the squared magnitude of the Fourier transform of the modulated unknown at the detectors,
\begin{equation}\label{eq:cdp}
    y_m = \left|\sum_{n=0}^{N-1}x[n]d_{l}[n]e^{-j2\pi kn/N}\right|^2, 
\end{equation}
where $d_l$ denotes the $l$-th modulating code. Here, $m = (l,k)$ where $0 \leq k \leq N-1$, and $1\leq l\leq L$, $L$ denoting the total number of modulating codes. Thus, by changing $d_l$, we generated different patterns of the unknown. Moreover, the $d_l$'s were independently and identically drawn from a distribution, known as the \textit{octanary pattern}\cite{candes2015phase2}.

\subsubsection{Noise and Outlier Model}
We assume the noise is uniformly distributed, i.e., each entry of $\w$ is independently and identically drawn from the uniform distribution, $\mathcal{U}(0,w_{max})$, satisfying $w_{max} \leq \sigma \|\x\|_2^2$, for some $\sigma > 0$. In addition, we generated a sparse outlier vector which had elements identically and independently drawn from the distribution defined in \eqref{eq:outlier_pdf}. We varied the values of $\sigma$ and $\theta$ to generate different levels of additive noise and outliers, respectively.


\subsection{Results of the RKLD-WF algorithm}
We compared the performance of the RKLD-WF algorithm with that of the standard WF algorithm \cite{candes2015phase} using $\ell_2$-loss and Poisson-loss functions, and the RWF  algorithm \cite{zhang2016reshaped} using the reshaped $\ell_2$-loss function, in terms of reconstruction accuracy, convergence and sample efficiency. For this purpose, we used the codes\footnote{https://viterbi-web.usc.edu/soltanol/WFcode.html, https://github.com/hubevan/reshaped-Wirtinger-flow} provided with the original papers, and used the simulation parameters provided by the authors.

\subsubsection{Reconstruction Accuracy}
First, we assessed the quality of reconstruction by the RKLD-WF algorithm using the CDP model. 
We set $N=512$ and the number of modulating patterns to $L=5$. The step size, $\mu$, was chosen as $0.4$ and the regularization parameter was set to $\lambda = 10^{-8}$. We fixed the number of iterations to $K=1000$ and ran $100$ $MC$ simulations. We then computed the ARE defined in (\ref{eq:avg_err}) at each iteration. 
Figures \ref{fig:wfplot1}(a) shows ARE vs. iteration count using noise-free data. Here, we see that all the RKLD-WF algorithm converges to an ARE around $10^{-15}$, thereby achieving an empirical \emph{exact recovery} of the unknown; however, due to the small number of samples ($L=5$), the WF-$\ell_2$, WF-Poisson and RWF algorithms fail to converge in some \emph{MC} trials which leads to a higher ARE than RKLD-WF. Moreover, RKLD-WF has a faster convergence speed than the other algorithms.
The faster convergence may be attributed to RKLD-WF utilizing a constant step-size.
Conversely, the WF-$\ell_2$ and WF-Poisson algorithms require a variable step-size, $\mu_k = \text{min} \left(1- e^{-k/k_0}, \mu_{max}\right)$. Here $k_0$ and $\mu_{max}$ are two additional hyper-parameters that require tuning. We fixed, $k_0 = 330$ and $\mu_{max} = 0.2$ which is much smaller than the step-size employed by RKLD-WF.

\subsubsection{Sample Efficiency}
We also evaluated the sample efficiency of RKLD-WF and compared to that of WF-$\ell_2$, WF-Poisson and RWF algorithms. For this experiment, we fixed $N=256$ for the CDP model and swept $L$ over $[2, 8]$ with a step-size of $1$. For the Gaussian model, the signal length was set to $N=100$ and the oversampling factor $\alpha$ was varied between $\left[2,6\right]$ with a step-size of $0.1$. We ran $100$ \emph{MC} simulations with $K = 1000$ for each value of $\alpha$ and $L$. Figures \ref{fig:wfplot1}(b) and (c) show the empirical probability of success vs. the oversampling factors, for the CDP and Gaussian models, respectively. For the CDP case, RKLD-WF has a sample efficiency notably higher than those of the WF-$\ell_2$ and WF-Poisson algorithms. For the Gaussian case, it can be clearly seen that the required number of samples for RKLD-WF is significantly lower than those of the other algorithms implying a higher empirical sample efficiency of the RKLD-WF algorithm. 

\subsubsection{Robustness of RKLD-WF in the Presence of Outliers} 
We evaluated the RKLD-WF algorithm using the CDP model, in the presence of outliers, to demonstrate the natural robustness of the RKLD-based loss function. Here, we used a complex Gaussian signal such that $\|\x\|_2^2=1$, with $N=128$ and $L=8$. We fixed the outlier fraction, $\rho = 0.1$, $\theta_{min}=5$ and varied $\theta_{max}$ between $4$ and $20$ with a step-size of $2$. Using a fixed measurement model $\bA$ and random outlier vectors, we generated different sets of measurements for $50$ $MC$ simulations and both algorithms were provided with the same initial estimate. Figure \ref{fig:robust_WF_l2_kl} shows the ARE with respect to $\theta_{max}$ for the RKLD-WF, WF-$\ell_2$, WF-Poisson and RWF algorithms. It can be seen that on increasing the values of $\theta_{max}$, the ARE for RKLD-WF remains lower than the other algorithms, maintaining a consistent gap. This result implies that even in the presence outliers, RKLD-WF converges and is able to recover the unknown up to a certain small reconstruction accuracy, thereby demonstrating a natural robustness to outliers and supporting our study.

\subsection{Results of the RKLD-MTWF and RKLD-GTWF Algorithms in the Presence of Noise and Outliers}
We performed numerical experiments to evaluate the performance of the RKLD-MTWF and the RKLD-GTWF algorithms, in the presence of \emph{additive} noise and outliers. In these experiments, we used the CDP model and a real unknown, $\x \in \mathcal{N}(0,\mathbf{I})$. We compared the outcomes of our algorithms with those of the median-TWF\footnote{https://github.com/hubevan/Median-TWF} and median-RWF algorithms in \cite{zhang2016reshaped}.

\subsubsection{Sample Efficiency}
Figure \ref{fig:mtwfplot1}(a) shows the empirical probability of success with respect to the number of modulating patterns, $L$, to evaluate the sample efficiency for the CDP model. Here, we set the signal length to $N=128$. We fixed the outlier magnitude parameters as $\theta_{min} = 5$ and $\theta_{max}=10$ and the fraction of outliers to $\rho = 0.1$. Looking at Figure \ref{fig:mtwfplot1}(a), We see that the RKLD-MTWF and RKLD-GTWF algorithms requires smaller number of samples for the $100\%$ recovery than those of the median-TWF and median-RWF algorithms, demonstrating 
superior sample efficiency. 

\begin{figure}
    \centering
    \includegraphics[width =0.5\columnwidth]{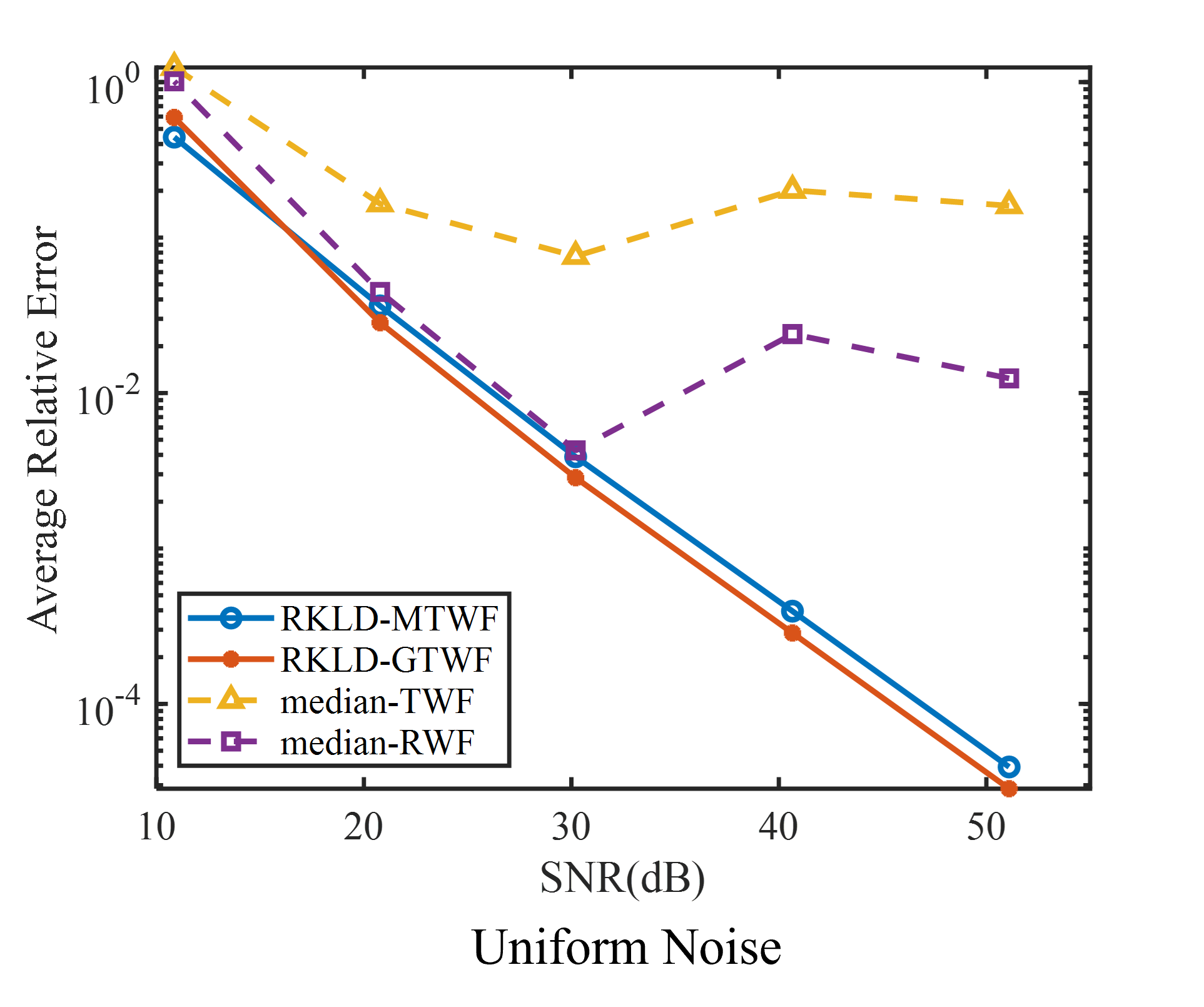}
    \caption{ARE vs. SNR (dB) for the CDP model in the presence of uniformly distributed additive noise and outliers. The different SNR levels are generated by varying the noise magnitude, $\sigma$.}
    \label{fig:snr1}
\end{figure}


\subsubsection{Reconstruction Accuracy}
To assess the performance of our algorithms in terms of reconstruction accuracy, we plotted the ARE with respect to $\theta_{max}$ in Figure \ref{fig:mtwfplot1}(b) where $\theta_{max} \in [0, 50]$ for fixed $\theta_{min}=0$, $\rho = 0.1$ and $\sigma = 0.01$. Here, the signal was $N=128$ and the number of modulating patterns were, $L=10$. We observe that our algorithms, RKLD-MTWF and RKLD-GTWF achieve a significantly lower ARE, on the order of $10^{-3}$, than those of the median-TWF and median-RWF algorithms\cite{zhang2016reshaped}. We also note that RKLD-GTWF obtains slightly lower error level than RKLD-MTWF. 

Additionally, we evaluated the outcomes of our algorithms in the presence of increasing percentage of outliers in the data\footnote{To eliminate the effect of noise, we only used data corrupted with outliers.}. Note that Figure \ref{fig:mtwfplot1}(c) shows the empirical probability of success (EPS) vs. the fraction of outliers, $\rho$. Here, we fixed the outlier parameters as $\theta_{min} =  5$ and $\theta_{max}=10$ and varied $\rho \in [0, 0.4]$. It can be  clearly seen that RKLD-GTWF and RKLD-MTWF can tolerate a considerably larger fraction of outliers in the data than the median-TWF and median-RWF algorithms, thereby exhibiting superior robustness against outliers to these algorithms.

Next, we assessed the reconstruction accuracy of our algorithms in the presence of only noise and fixed outliers. For this experiment, we set $N = 100$ and $L = 10$. The outlier parameters were fixed at $\theta_{min}=5$ and $\theta_{max}=10$. We generated the measurements at different levels of SNR by varying the noise magnitude parameter $\sigma$. We averaged the results of $100$ \emph{MC} simulations to generate the ARE. Figure \ref{fig:snr1} shows ARE vs. SNR (dB) for the Gaussian model. We see that the ARE decreases gracefully with increasing values of SNR for the RKLD-MTWF and RKLD-GTWF algorithms whereas the median-TWF and median-RWF yield higher ARE at different SNR levels.


\section{Performance Evaluation Using Real Imaging Dataset}\label{sec:real_data}
We evaluated the performance of our RKLD-WF algorithm for phaseless imaging using the real measurements provided in \cite{metzler2017coherent}. In addition, we compare the reconstructed images of our algorithm to the baseline reconstructions of the WF-$\ell_2$, TWF, and RWF algorithms. To reproduce the images by the baseline algorithms, we used the corresponding modules provided in the \emph{PhasePack} library \cite{chandra2019phasepack}.

\subsection{Real Imaging Data}
\begin{figure}[!h]
     \centering
     \includegraphics[width=0.6\columnwidth]{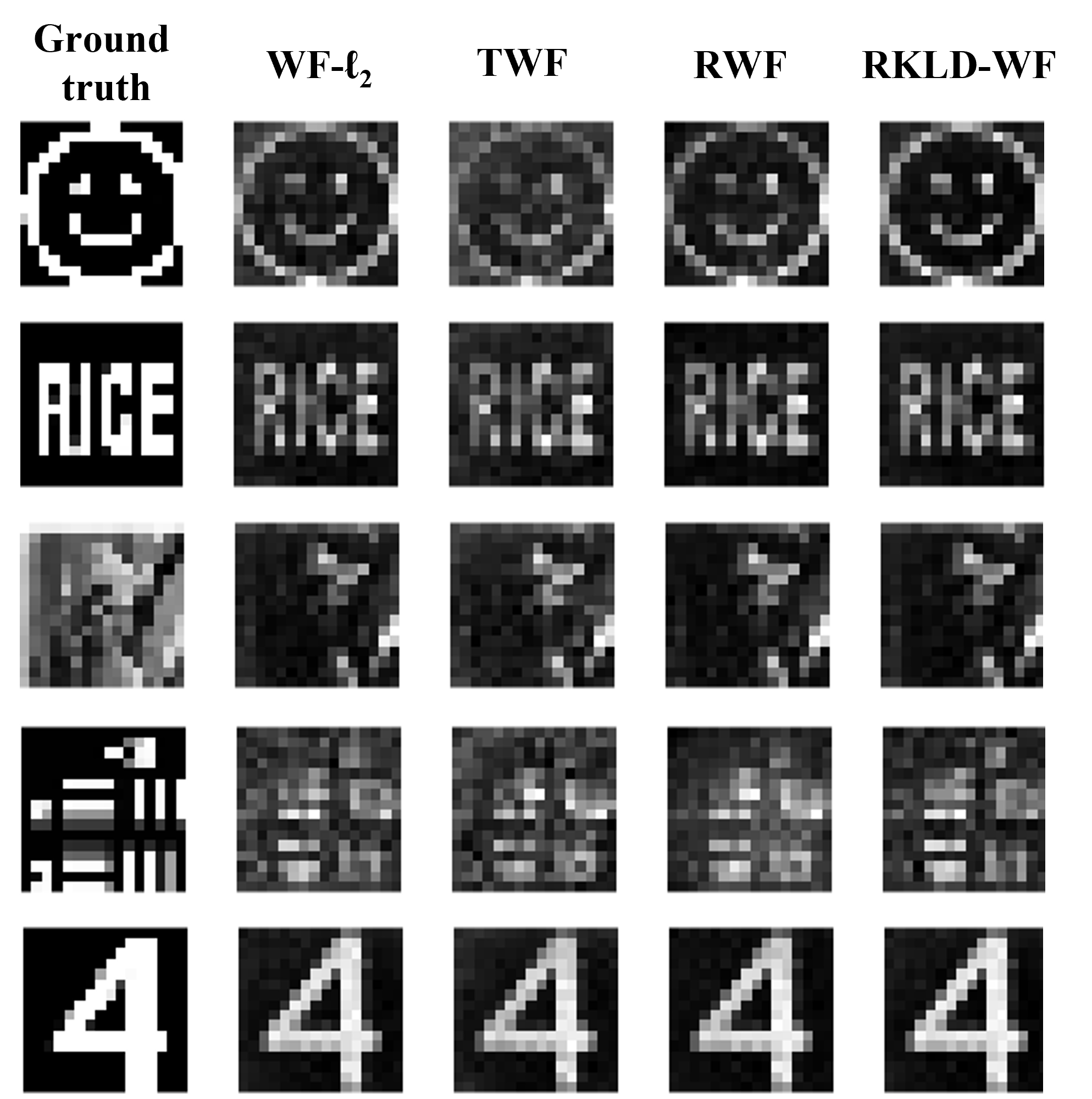}

     \caption{Comparison of the ground truth (left), and the average absolute images reconstructed via WF-$\ell_2$, TWF, RWF, and our RKLD-WF (right), respectively. No outliers added to the data. All methods are optimized with backtracking line-search, iterations stopped at the identical number of iterations required by the RKLD-WF method.}
    \label{fig:img0}
\end{figure}

The setup employs a spatial light modulator which modulates a coherent light using five ground-truth imaging patterns shown in the first column of Figure \ref{fig:img0}. The modulated light  passes through a diffuser and lands on a sensor that measures only the intensity values. In this setup,  $\bA$ characterizes the linear mapping from the image pattern to the measurements as light passes through the scattering medium. In this experiment, we used the $\bA$ matrix provided in \cite{metzler2017coherent}.

\begin{figure}[!h]
     \centering
     \includegraphics[width=0.6\columnwidth]{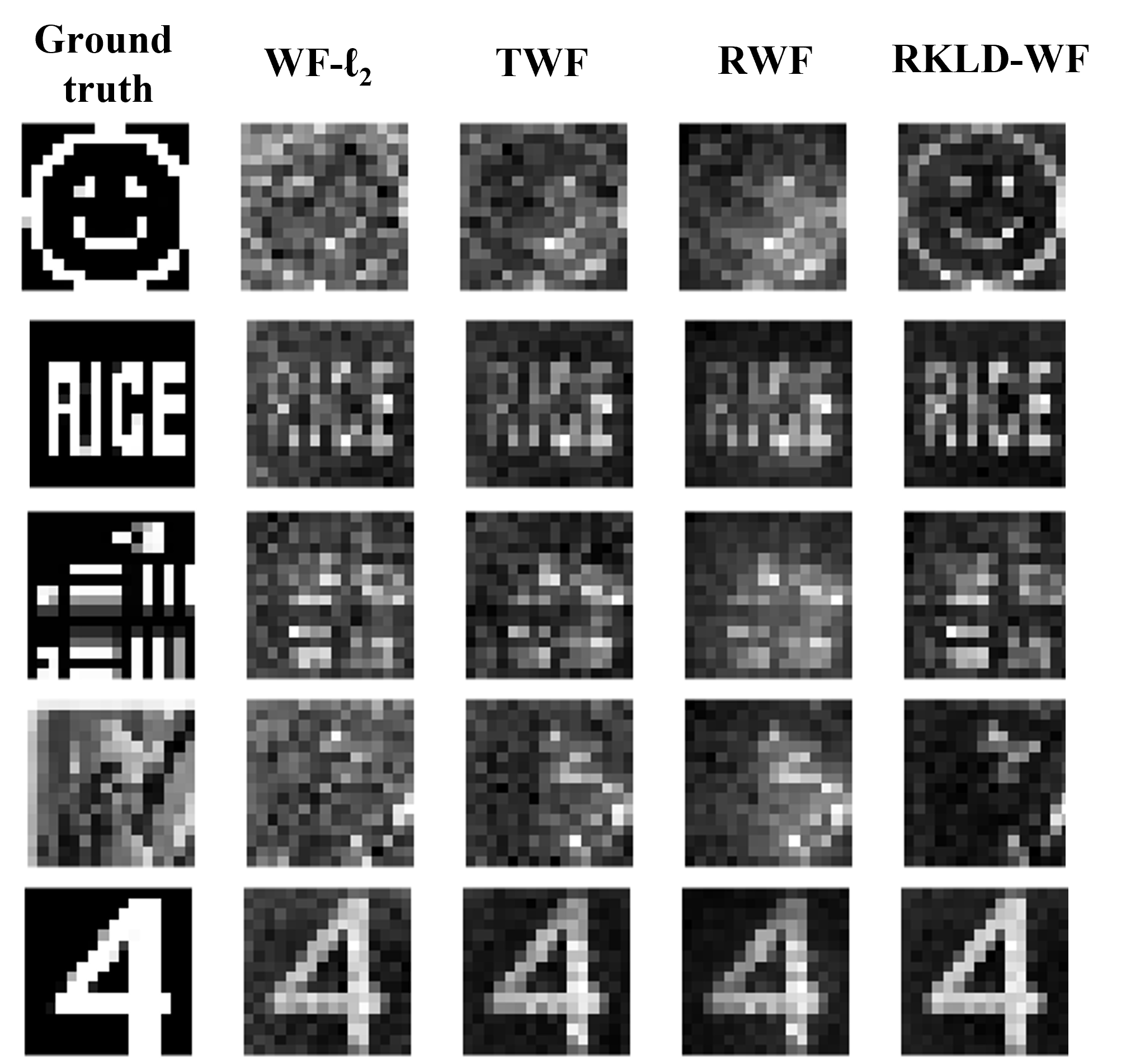}

     \caption{Comparison of the ground truth (left), and the average absolute images reconstructed via WF-$\ell_2$, TWF, RWF, and our RKLD-WF (right), respectively. Outliers added to the data with magnitude parameter, $\theta_{min} =1$ and $\theta_{max}=2$. No truncation is used on RKLD-WF and WF-$\ell_2$. All methods are optimized with backtracking line-search, iterations stopped at the identical number of iterations required by the RKLD-WF method.}
    \label{fig:img1}
\end{figure}

\begin{figure*}[!ht]
\centering
  \subfloat[\label{1b}]{%
        \includegraphics[width = 0.45\columnwidth]{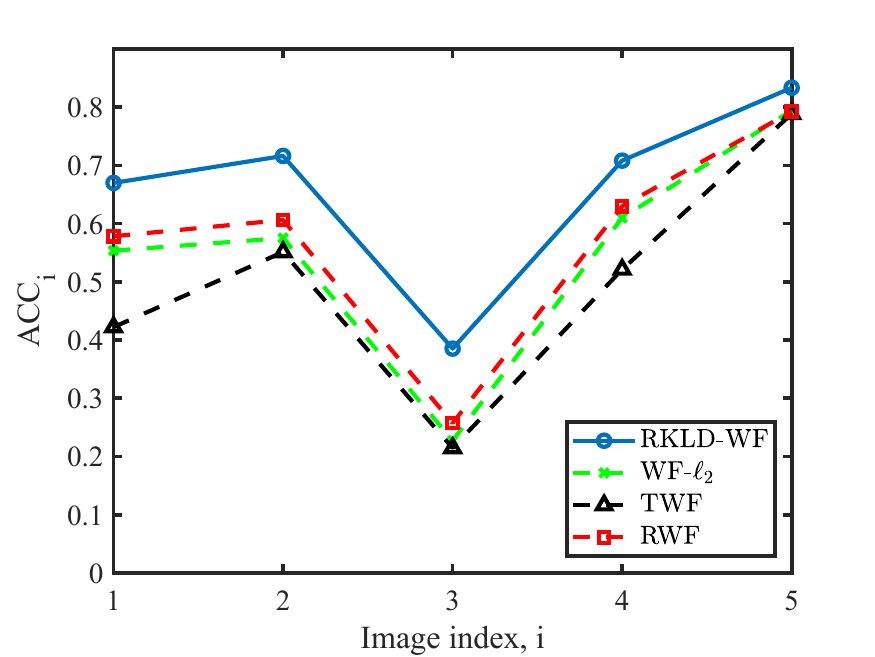}}
        \hspace{0.1cm}
    \subfloat[\label{1a}]{%
       \includegraphics[width = 0.45\columnwidth]{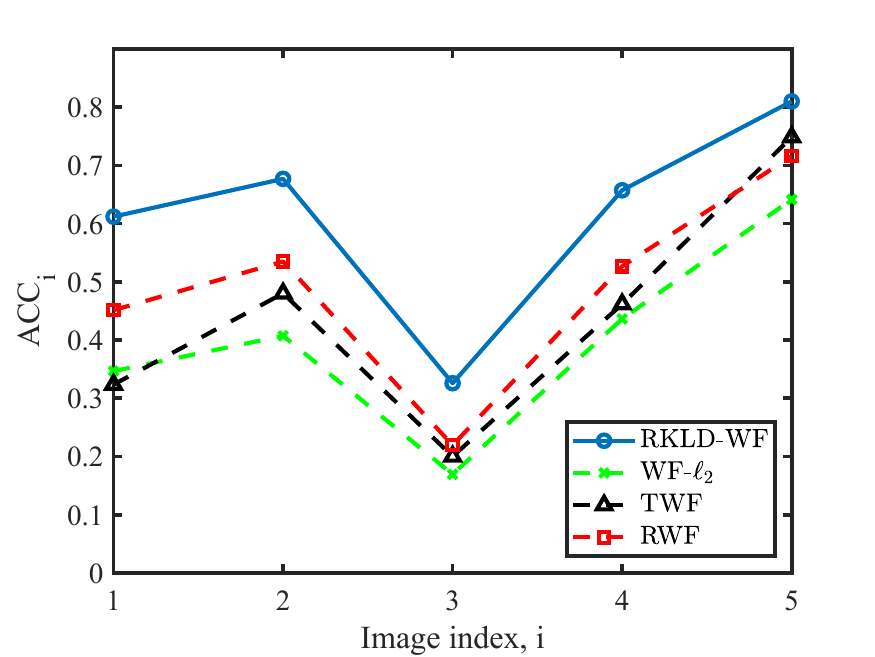}}
       
    \caption{\emph{(a)} Average correlation coefficient over the $50$ Monte-Carlo instances with no added outlier, at $\alpha = 5$, for each of the $5$ test images. \emph{(b)} Average correlation coefficient over the $50$ Monte-Carlo instances with outlier with $\theta_{min}=1$ and $\theta_{max}=2$, at $\alpha = 5$, for each of the $5$-test images.}
    \label{fig:my_label}
\end{figure*}
We consider the $16\times 16$ reconstruction task for our simulation, where the measurements have an estimated SNR of $22$dB. The full dataset contains $M = \alpha N =256^2$ measurements for each image, where the signal length is $N = 256$ and the oversampling factor is $\alpha = 256$. For our experiment, we used randomly sampled subsets of measurements from the full dataset at varying oversampling factors.

\subsection{Simulation Set-up}
We performed our experiments both with and without the additive outliers. To assess the impact of outliers, we added positive outliers to the measurements using the distribution in \eqref{eq:outlier_pdf}. We fixed $\theta_{min}=1$ and $\theta_{max}=2$ and set the outlier fraction to $\rho = 0.1$. We ran $50$ \emph{MC} simulations where, for each simulation, different subsets of measurements were used by randomly sampling from the full data-set. To maintain consistency, all the other algorithms with whom RKLD-WF is compared to are run for the same number of iterations required by the RKLD-WF to converge.

\subsection{Performance Evaluation Metrics}
To assess the performance of the algorithms, we evaluated the quality of the reconstructed images via the average correlation coefficient ($\text{ACC}_i$) using $S$ $MC$ simulations, i.e.,
\begin{equation}\label{eq:CC2}
\text{ACC}_i( \x , \hat{\x}) := \frac{1}{S}\sum_{s=1}^{S}\frac{| \langle \x_i , \hat{\x}^s_i \rangle |}{\| \x_i \|_2}
\end{equation}
where  $\hat{\x}^s_i$ is the unit norm reconstruction of $\x_i$, at the $s$-th MC simulation and for the $i$-th image, where $i \in \{1, 2, 3, 4, 5\}$ denotes the indices of the $5$ different images in the real imaging data-set\cite{metzler2017coherent} shown in the first column of Figure \ref{fig:img1}. Additionally, we evaluate the empirical sample efficiency for which we use the average of $\text{ACC}_i$'s over the $5$ images:
\begin{equation}\label{eq:CC2}
\text{ACC}_{\alpha}( \x , \hat{\x}) := \frac{1}{5}\sum_{i=1}^{5} \text{ACC}_i(\x, \hat{\x}).
\end{equation}

\subsection{Results and Comparative Analysis of Different Algorithms}
First, we performed reconstruction using the data free of outliers. 
Figure \ref{fig:img0} shows the average absolute value images obtained from the MC experiments at $\alpha = 5$. Furthermore, Figure \ref{fig:img1} shows the average images reconstructed using the data corrupted by outliers. The corresponding $\text{ACC}_i$ vs. image index are shown in Figures \ref{fig:my_label}(a) and \ref{fig:my_label}(b) for the cases of without and with the outliers, respectively. 

Most strikingly, our RKLD-WF method does not employ a truncation scheme in the presence of outliers. In contrast, the Poisson-loss incorporates a truncation mechanism (the TWF algorithm \cite{chen2017solving}) to provide robustness. Although truncation and reshaping provide improvements over the standard WF-$\ell_2$ in the presence of outliers, RKLD-WF shows superior reconstruction performance at identical total computational complexity without any truncation. This result highlights the natural robustness of the RKLD-based  minimization , which supports our analysis in Section $3$ .

\begin{figure}[!h]
     \centering
     \includegraphics[width =0.6\columnwidth]{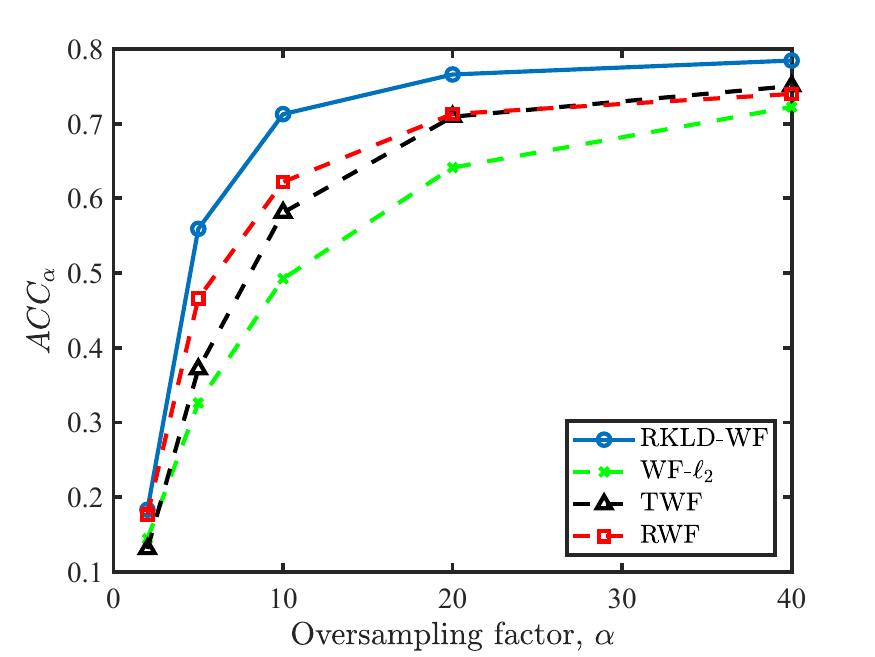}
     \caption{Average correlation coefficient ($\text{ACC}_\alpha$) vs. the oversampling factor over the $5$ images in the optical imaging dataset \cite{metzler2017coherent}. We show the results for the RKLD-WF, WF-$\ell_2$ TWF, and RWF algorithms using $50$ MC simulations for each choice $\alpha$.}
    \label{fig:CC}
\end{figure}

Finally, Figure \ref{fig:CC} shows the $\text{ACC}_{\alpha}$  vs. the oversampling factor $\alpha$ for all four algorithms to demonstrate the sample efficiency. Here, we performed reconstructions by randomly sampling measurements for different choices of $\alpha \in \{5, 10, 20, 25, 40\}$. It can be seen that the RKLD-WF algorithm has a higher correlation coefficient than those of the other three algorithms for smaller values of the oversampling factor. This observation shows that RKLD-WF provides superior recovery in sample-starved regions.  

Table \ref{tab:iter} tabulates the iteration counts demonstrating the convergence speed of algorithms. The number of iterations was averaged over the images and recorded for each value of $\alpha$. All four algorithms ran for a maximum of $1000$ iterations. Here, an early stopping criterion was implemented: if the relative error between the subsequent iterates fell below $10^{-10}$ for three consecutive iterations, the iterations would terminate. 
We see that the RKLD-WF converges much faster than the other three algorithms. This result also agrees with our arguments and illustrates the advantage of the RKLD-based optimization.

\begin{table}[!h]
\caption{The average iteration count over $50$ \emph{MC} simulations required by the RKLD-WF, WF-$\ell_2$, TWF, and RWF algorithms for oversampling fators $\alpha \in \{2, 5, 10, 20, 40\}.$}
\centering
\begin{tabular}{c|ccccc|}
\cline{2-6}
\multicolumn{1}{l|}{\multirow{2}{*}{}} & \multicolumn{5}{c|}{oversampling Factor, $\alpha$}                                                                             \\ \cline{2-6} 
\multicolumn{1}{l|}{}                  & \multicolumn{1}{c|}{2}      & \multicolumn{1}{c|}{5}      & \multicolumn{1}{c|}{10}     & \multicolumn{1}{c|}{20}     & 40     \\ \hline
\multicolumn{1}{|c|}{RKLD-WF}           & \multicolumn{1}{c|}{311.1}    & \multicolumn{1}{c|}{27.8}   & \multicolumn{1}{c|}{14.2}   & \multicolumn{1}{c|}{11.7}     & 8.7    \\ \hline
\multicolumn{1}{|c|}{WF-$\ell_2$}            & \multicolumn{1}{c|}{1000}   & \multicolumn{1}{c|}{651.6}  & \multicolumn{1}{c|}{187.6}  & \multicolumn{1}{c|}{103.2}  & 88.2  \\ \hline
\multicolumn{1}{|c|}{TWF}              & \multicolumn{1}{c|}{987.7} & \multicolumn{1}{c|}{881.2} & \multicolumn{1}{c|}{745.8}   & \multicolumn{1}{c|}{760.9} & 779.3  \\ \hline
\multicolumn{1}{|c|}{RWF}              & \multicolumn{1}{c|}{1000}   & \multicolumn{1}{c|}{1000}   & \multicolumn{1}{c|}{1000} & \multicolumn{1}{c|}{997.7} & 937.8 \\ \hline
\end{tabular}
\label{tab:iter}
\end{table}

\section{Conclusion}\label{sec:conclusion}
In this paper, we approach the phase retrieval from a novel perspective and present an RKLD-based optimization method to attain robustness against additive outliers beyond a level offered by state-of-the-art algorithms. 
The robust inference is crucial for the practical applicability of the phase retrieval algorithms as measurements are commonly collected under model imperfections and challenging acquisition conditions which can considerably degrade the quality of measurements. 
The RKLD provides a robust measure of dissimilarity for processing intensity-only data and promotes a model agnostic approach in formulating the optimization problem. 
We presented a robustness analysis of the RKLD-based loss function and compared it to the robustness of $l_2$ and Poisson loss functions. Additionally, we presented three RKLD-based algorithms and assessed their performance as compared to state-of-the-art robust phase retrieval algorithms using synthetic and real phaseless measurements. Our extensive numerical results support our theoretical analysis and demonstrate the superiority of our algorithms.

Our future work will  focus on the theoretical convergence analysis of the RKLD-based optimization  to establish recovery guarantees. 

\section{Acknowledgments}
This  work  was  supported  in  part  by  the  Air  Force  Office  of  Scientific Research (AFOSR) under the agreement FA9550-19-1-0284, in part by Office of Naval Research (ONR) under the agreement N0001418-1-2068, in part by the  United  States  Naval  Research  Laboratory  (NRL)  under  the  agreement N00173-21-1-G007,  and  in  part  by  the  National  Science  Foundation  (NSF) under Grant ECCS-1809234.

\bibliographystyle{IEEEtran}
\bibliography{Bib_alt, Bib_v2, Bibliography, IEEEabrv}








\end{document}